# Employing long fibre-optical Mach-Zehnder interferometers for quantum cryptography with orthogonal states

G. B. Xavier, G. P. Temporão and J. P. von der Weid

We experimentally demonstrate that a long distance actively stabilized Mach-Zehnder fibre optical interferometer can be used to reliably implement the GV95 quantum cryptography protocol employing orthogonal states. A proof-of-principle experiment using an interferometer composed of 1 km of spooled optical fibres connecting Alice and Bob is performed. The active stabilisation against phase drifts is done with a classical channel wavelength-multiplexed with the quantum channel and a high stable visibility implying a QBER of 2.2 % is shown.

*Introduction:* Quantum key distribution (QKD) [1] is a theoretically proven way to securely distribute random keys between two distant communicating parties. Most QKD protocols are based in the Bennett-Brassard protocol (BB84) [1]. In this approach, two bases (pairs of orthogonal states) are used for encoding the random bits, such that any two states from a different basis are always non-orthogonal, which guarantees the security of the protocol. However, in 1995, Goldberg and Vaidman proposed a QKD protocol in which just two orthogonal quantum states, composed of superpositions of localized wave packets, are used to distribute a secret key between Alice and Bob [2]. They have shown that if the wave packets are not simultaneously sent by Alice, the delay $\tau$ between them is larger than the uncertainty in emission and measurement times, and all emission times are random, then the security of the protocol is guaranteed. Although having been proposed for quite some time, the GV95 protocol was only very recently experimentally demonstrated in a proof-of-principle experiment on an optical table employing bulk optics [3].

In order to have real practical applicability, a demonstration is needed over longer distances. In this work we demonstrate the feasibility of the GV95 protocol over a 1 km long actively phase-stabilized Mach-Zehnder (MZ) interferometer composed of telecom single-mode optical fibres [4]. All the components used in the demonstration are commercial ones, enhancing the feasibility of implementing the GV95 protocol in real telecom systems.

*Experimental setup:* The two orthogonal states required by the protocol are $|\psi_0\rangle = 1/\sqrt{2}(|a\rangle + |b\rangle)$ and $|\psi_1\rangle = 1/\sqrt{2}(|a\rangle - |b\rangle)$ where $|a\rangle$ and $|b\rangle$ are localized wavepackets in the spatial modes at the output of the first beamsplitter (BS) of the MZ interferometer as shown in Fig. 1. The two states, $|\psi_0\rangle$ and $|\psi_1\rangle$, can be produced depending on which input port of the interferometer the single-photon is sent, $S_0$ and $S_1$, with each state manually chosen with the optical switch (OS).

In this proof-of-principle experiment the single photons are produced from a distributed feedback (DFB) semiconductor laser working under continuous-wave (CW) mode, followed by an optical attenuator. The wavelength of the single photons is $\lambda_Q = 1546.12$ nm, and the measured coherence length is $\sim 6.4$ m. The attenuator is set such that there is an average of 0.1 photon / detection window at each input port of the MZ interferometer. One of the outputs of the optical switch is connected directly to the 50:50 coupler, while the other is first connected to a dense wavelength division multiplexer (DWDM), to be combined with the interferometer control signal. This signal for the active phase stabilisation system is generated from an external cavity tuneable laser, centred at $\lambda_{PH} = 1547.72$ nm with $> 50$ m coherence length. The DWDM has a 1.6 dB insertion loss, and as explained in [4], its use combined with the Bragg gratings is enough to filter out noise photons from the control channel. Manual polarisation controllers are used after the optical switch to ensure that both $S_0$ and $S_1$ modes arrive at the input coupler with the same state. Another polarisation controller is used to adjust the polarisation state of the control laser, such that it is the same as the single photons entering the interferometer, aiding in the alignment of the setup.

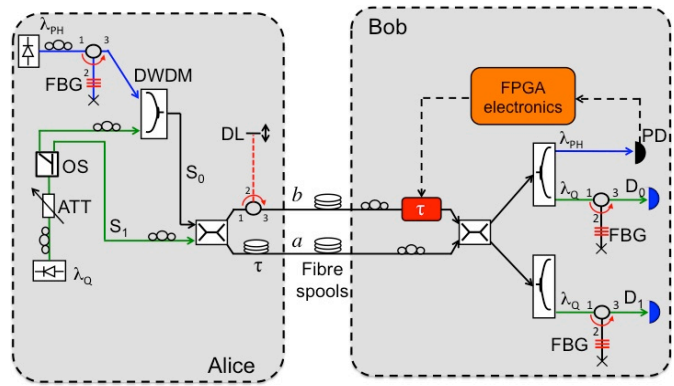

**Fig. 1** *Experimental setup. ATT: optical attenuator; DL: optical delay line; DWDM: dense wavelength division multiplexer; FBG: fibre Bragg grating; FPGA: field programmable gate array; OS: optical switch; PD: p-i-n photodetector. The delay $\tau$ located at Bob's station in mode b is implemented by the same fibre stretcher used to keep the interferometer locked in phase.*

Following the 50:50 input BS, there is an optical delay line in mode *b* comprised of an optical circulator, a free-space fibre coupler with a collimator and a mirror mounted on a linear translation stage (total loss $\sim 3$ dB). This delay line is manually used, together with a filtered broadband light source, to coarsely adjust the length mismatch of the two interferometer arms to be within $\sim 1$ mm. In mode *a* the delay $\tau$ is implemented with a short spool of 40 m of optical fibre.

The interferometer is composed of 1 km of optical fibres contained in the same spooled cable in the lab. The output of the cable is connected to Bob's site, which contains a fibre-optical piezo-electric stretcher in mode *b*. The length of the fibre stretcher is 40 m, and it is used as the optical delay $\tau$ needed at Bob. Following the upper output port of the interferometer in Bob's station, the quantum and classical channels are separated with a DWDM, identical in operation to the one used in Alice. The classical signal is detected by a p-i-n photodetector (PD), and its output read by the control electronics, based on a field programmable gate array (FPGA). The control algorithm acts on the fibre stretcher to keep the interferometer free from phase fluctuations [4].

The pseudo-single photons are detected by commercial InGaAs-based single-photon counting modules (SPCM) $D_0$ and $D_1$, working under gated mode, at 500 kHz repetition frequency, 2.5 ns wide gate window and dark count probabilities per gate of $1.4 \times 10^{-5}$ and $3.87 \times 10^{-5}$ respectively. Before detection at $D_0$ and $D_1$, another combination of FBGs and circulators are used to filter out any crosstalk produced in the DWDM. Manual polarisation controllers inside the interferometer are used to adjust the visibility of the interference fringes to be maximum at the detectors.

*Results and discussions:* With the interferometer adjusted such that there is a minimum probability of detection at $D_0$ ($|\psi_1\rangle$ is sent), the photon counts per second in both $D_0$ and $D_1$ are simultaneously acquired. After a short period of time, the optical switch is flicked, such that the state $|\psi_0\rangle$ is sent to Bob. The counts are acquired once more for a new period of time, the switch is manually changed again and so on for the remainder of the measurement. These results are shown in Fig. 2. It should be stressed that the phase control system was not restarted or readjusted during the entire experimental run.

One can observe that when state $|\psi_0\rangle$ is sent (minimum probability of a photon arriving at $D_1$), the counts are higher than at $D_0$ for state $|\psi_1\rangle$. This is due to the considerably higher dark counts generated from $D_1$. The integration time for each acquired data point shown in Fig. 2 is 1 s.

We evaluate the performance of the experiment by calculating the optical visibility and the quantum bit error rate (QBER), obtained from the data in Fig. 2. First the visibility is calculated by grouping together the results for both states $|\psi_0\rangle$ and $|\psi_1\rangle$. It is then calculated for each transmitted state as $V = (C_0 - C_1)/(C_0 + C_1)$, where $C_0$ are the counts per unit of time in detector $D_0$ and $C_1$ in detector $D_1$. The raw visibilities (without subtracting dark counts) for $|\psi_0\rangle$ and $|\psi_1\rangle$ are $0.902 \pm 0.011$ and $0.962 \pm 0.009$. Calculating the net (removing the dark counts) visibility these values rise to $0.978 \pm 0.012$ and $0.989 \pm 0.009$, demonstrating indeed that the higher dark counts of $D_1$ are responsible for the higher noise level shown in Fig. 2 by the same detector.

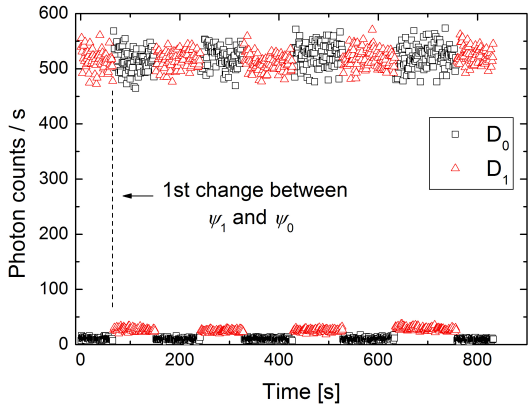

**Fig. 2** *Raw photon counts at both detectors $D_0$ and $D_1$ over a period of approximately 14 minutes. The optical switch is changed a few times over the course of the experiment to generate either $|\psi_0\rangle$ or $|\psi_1\rangle$. A dashed vertical line indicates the first change of the switch. Note that the phase drift is kept stable independently of which state is sent, allowing the protocol to be successfully executed.*

We calculate the error rate from its definition: $QBER = (N_{right} - N_{wrong})/(N_{right} + N_{wrong})$, where $N_{wrong}$ and $N_{right}$ are the number of incorrect and correct received bits respectively, detected by Bob per integration time. For example, in the case of state $|\psi_0\rangle$ the correct bits are the counts received at $D_1$ and the incorrect ones the counts detected at $D_0$. The raw QBER for the transmitted states $|\psi_0\rangle$ and $|\psi_1\rangle$ is $4.88 \pm 0.56\%$ and $1.91 \pm 0.45\%$, once again reflecting the fact that the dark count rate of $D_1$ is higher. Recalculating the QBER by correcting the counts produced by $D_1$ assuming as if it has the same dark count rate of $D_0$, we obtain an average total QBER for the experiment, which is $2.21 \pm 0.62\%$. We would like to note that the employed delay $\tau$ of 40 m, which is equivalent to a propagation time of ~ 192 ns, is greater than the uncertainty in the emission time (~ 30.7 ns coherence time) and in the detection (2.5 ns gate window), as required by GV95 [4].

*Conclusion:* We have successfully demonstrated that the GV95 protocol is feasible over 1 km long optical fibres, with an inferred QBER comparable to many other QKD setups. In addition, it should be possible to extend the 1 km range to a distance of a dozen kms, by employing a faster piezo-electric actuator. As the GV95 protocol is experimentally demonstrated, a need for a thorough theoretical security analysis becomes crucial in order to determine its applicability in a real-world scenario.

*Acknowledgments:* G. B. X. acknowledges financial support from Milenio project P10-030-F, CONICYT PFB08-024 and FONDECYT no. 11110115; G. P. T and J. P. W. acknowledge financial support from CAPES, FAPERJ and CNPq.

G. B. Xavier (Departamento de Ingeniería Eléctrica, Universidad de Concepción, Casilla 160-C, Concepción Chile).

E-mail: gxavier@udec.cl

G. P. Temporão and J. P. von der Weid (Centre for Telecommunication Studies, Pontifical Catholic University of Rio de Janeiro, Rua Marquês de São Vicente, 225 Gávea – Rio de Janeiro - Brazil).

G. B. Xavier is also with MSI-Nucleus on Advanced Optics and Center for Optics and Photonics, Universidad de Concepción, Casilla 160-C, Concepción Chile.

**References**

1. Gisin N., Ribordy G., Tittel W. and Zbinden H., 'Quantum Cryptography', Rev. Mod. Phys., 2002, **74**, pp.145.

2. Goldenberg L. and Vaidman L., 'Quantum cryptography based on orthogonal states', Phys. Rev. Lett., 1995, **75**, pp. 1239.

3. Avella A., Brida G. Degiovanni I. P., Genovese M., Gramegna M. and Traina P., 'Experimental quantum-cryptography scheme based on orthogonal states', Phys. Rev. A 2010, **82**, pp. 062309.

4. Xavier G. B. and von der weid J. P., 'Stable single-photon interference in a 1 km fiber-optic Mach-Zehnder interferometer with continuous phase adjustment', Opt. Lett. 2011, **36**, pp. 1764.